\newcommand{\un}{~\mathrm} 
\newcommand{\ie}{{\em i.e. }} 
\newcommand{\eg}{{\em e.g. }} 

\documentclass[namedreferences]{kluwer}

\usepackage{graphicx}
\usepackage{array}
\usepackage{dcolumn}
\usepackage{bm}
\usepackage{klucite}

\begin{document}
\begin{article}

\begin{opening}
\title{Nanoscale Damage During Fracture in Silica Glass}

\author{D. \surname{Bonamy}$^1$\email{bonamy@drecam.cea.fr}, S. \surname{Prades}$^{1,2}$, C.~L. \surname{Rountree}$^{1,3}$, L. \surname{Ponson}$^1$, D. \surname{Dalmas}$^{1,4}$, E. \surname{Bouchaud}$^1$, K. \surname{Ravi-Chandar}$^5$, C. \surname{Guillot}$^1$}
\institute{
$^1$ Fracture Group, Service de Physique et Chimie des Surfaces et Interfaces, DSM/DRECAM/SPCSI, CEA Saclay, F-91191 Gif sur Yvette, France\\
$^2$ Present address: Department of Materials, Swiss Federal Institute of Technology,
 ETH H\"{o}nggerberg, Wolfgang-Pauli-Strasse 10 CH-8093 Z\"{u}rich, Switzerland\\
$^3$ Collaboratory for Advanced Computing and Simulations$,$ Departments of Material Science \& Engineering$,$ Physics \& Astronomy$,$ Computer Science \& Biomedical Engineering University of Southern California, Los Angeles, CA 90089-0242, USA\\
$^4$ Present address: Glass Surface and Interface, Unit\'e Mixte CNRS/Saint-Gobain, F-93303 Aubervilliers Cedex, France\\
$^5$ Center for Mechanics of Solids, Structures and Materials, Department of Aerospace Engineering and Engineering Mechanics, University of Texas, Austin, TX 78712, USA}                                                              

\begin{abstract}
We report here atomic force microscopy experiments designed to uncover the nature of failure mechanisms occuring within the process zone at the tip of a crack propagating into a silica glass specimen under stress corrosion. The crack propagates through the growth and coalescence of nanoscale damage spots. This cavitation process is shown to be the key mechanism responsible for damage spreading within the process zone. The possible origin of the nucleation of cavities, as well as the implications on the selection of both the cavity size at coalescence and the process zone extension are finally discussed.
\end{abstract}


\abbreviations{\abbrev{KAP}{Kluwer Academic Publishers};
   \abbrev{compuscript}{Electronically submitted article}}

\nomenclature{\nomen{KAP}{Kluwer Academic Publishers};
   \nomen{compuscript}{Electronically submitted article}}

\classification{JEL codes}{D24, L60, 047}
\end{opening}

\section{Introduction}\label{intro}

Slicate glasses are often considered as the archetype of brittle materials. It is commonly thought (\opencite{Kelly67_pm}; \opencite{Lawn80_jms}; \opencite{Guin05_prl}) that its fracture is similar to cleavage -~with a crack progressing through sequential bond ruptures without involving any damage ahead of the crack tip. However, some recent observations calls into questions this scenario:

\begin{itemize}
\item The morphology of fracture surfaces in glasses exhibits scaling features similar to the ones observed in a wide range of quasi-brittle and ductile materials, \eg oxide glass, polymers, metallic alloys, wood, rocks... (see \inlinecite{Bouchaud97_jpcm} and references therein). This strongly suggests the existence of some underlying generic mechanisms within the process zone common to all these materials.  
\item The deformation field was shown not to fit with the linear elastic predictions over a fairly large region (of the order of a hundred of nanometers) in the vicinity of the crack tip (\opencite{Guilloteau96_epl}; \opencite{Heneau00_jacs}).
\item Molecular Dynamic (MD) simulations indicate that fracture in silica glass proceeds through the growth and coalescence of nanoscale cavities (\opencite{vanBrutzel02_mrssp}; \opencite{Rountree02_armr}; \opencite{Kalia03_ijf}). 
\end{itemize}

These various results have led us to investigate experimentally the failure mechanisms occuring in glass at its microstructure scale, the nanoscale. The first series of experiments, reported in \citeauthor{Celarie03_prl} \shortcite{Celarie03_prl,Celarie03_ass} and \inlinecite{Marliere03_jpcm}, were carried out on aluminosilicate vitroceramics. They clearly reveal that crack progresses through the growth and coalescence of nanoscale damage cavities. We describe here our recent studies performed on a minimal elastic vitreous medium, pure silica. 
The experimental setup is described in section 2. In section 3, we present the experimental results. As in the aluminosilicate vitroceramics, crack propagation proceeds through the nucleation, growth and coalescence of damage cavities (section 3.1). The rate of cavity nucleation and the size of cavities at coalescence are shown to set the mean crack growth velocity as measured at the continuous scale (section 3.2). This nanocavitation process is shown to set the process zone size (section 3.3). Finally, section 4 is devoted to a discussion on a possible scenario explaining the existence of damage cavities ahead of the crack tip, and its implications as regard to the cavity size at coalescence (section 4.1) and the process zone size(section 4.2).

\section{Experimental Setup}\label{sec_setup}

\begin{figure}[!hbtp]
\centering
\includegraphics[width=0.9\textwidth]{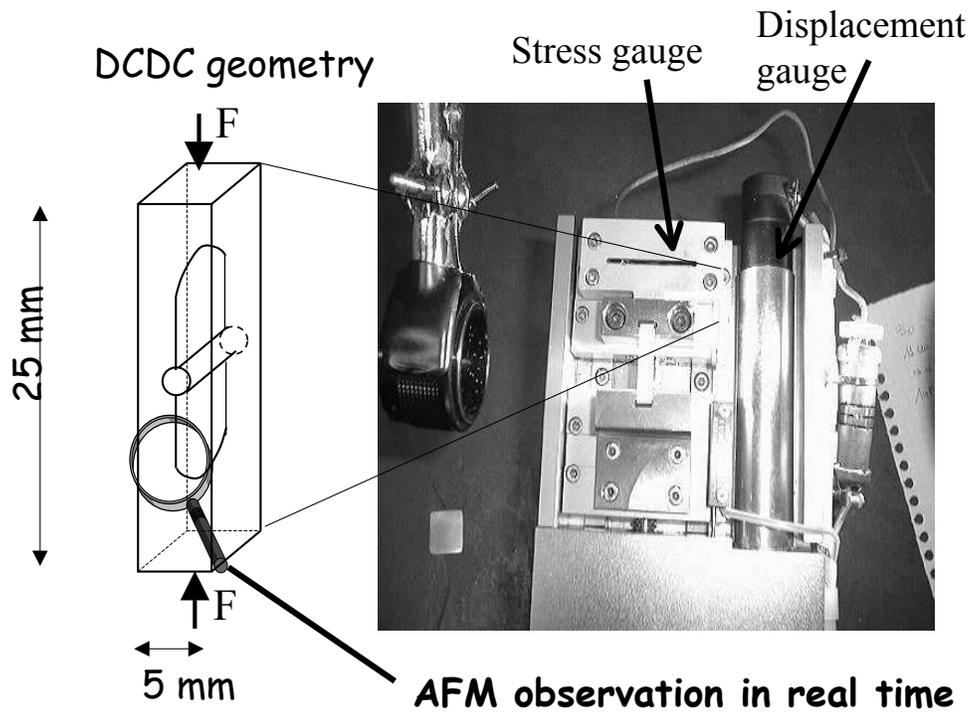}
\caption{Experimental setup showing the loading DCDC configuration used to fracture the glass specimen under stress corrosion.}
\label{setup}
\end{figure}

The experimental set-up has been described in detail in \inlinecite{Prades05_ijss} and is briefly recalled below. Fracture was performed on double cleavage drilled compression parallelepipedic samples (size $5\times5\times25 \un{mm}^3$) with a cylindrical hole drilled in the center (radius $0.5\un{mm}$). A gradually increasing uniaxial compressive load was applied to the sample (Figure \ref{setup}). Once the two cracks are initiated symmetrically from the hole, the load is held constant. In this geometry, the stress intensity factor $K_I$ can be related to the crack length using the expression given by \inlinecite{He95_amm}. While the cracks length increases, $K_I$ decreases. Under vacuum, the crack would stop when $K_I$ gets smaller than the material fracture toughness $K_{Ic}$. But, in our room atmosphere (relative humidity $45\%$, temperature $26^\circ \mathrm{C}$), the corrosive action of water on glass allowed slow, sub-critical, crack propagation. 
The crack tip advance was then slow enough to be probed by AFM.

\begin{figure}[!hbtp]
\centering
\includegraphics[width=0.6\textwidth]{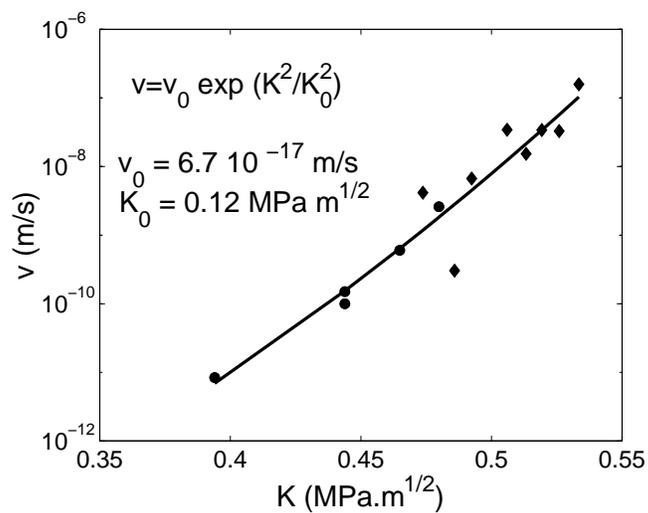}
\caption{Variation of the mean crack growth velocity $v$ (in $\un{m.s}^{-1}$) as a function of the stress intensity factor $K_I$ (in $\un{MPa.m}^{1/2}$). The axes are semi-logarithmic. Diamonds and dots correspond to optical and AFM measurements respectively. The line is a fit using $v =v_0\exp(K_I^2/K_0^2)$ as expected for stress enhanced activated process models.}
\label{v2K}
\end{figure}

The crack growth velocities $v$ considered in this study range between $10^{-12}\un{m/s}$ and $10^{-6} \un{m/s}$. In this velocity range, $v$ increases exponentially with $K_I^2$ (Figure \ref{v2K}) in agreement with stress enhanced activated process models as proposed by \inlinecite{Wiederhorn67_jacs} and \inlinecite{Wiederhorn70_jacs}. The velocity could then be tuned by adjusting the external applied load for a measured crack length. It was varied over three decades, from $10^{-9}$ and $10^{-12}\un{m/s}$ (the maximum reachable velocity being set by the recording time of an AFM frame, around 3 minutes).

In all the following, the reference frame $(\vec{e}_x,\vec{e}_y,\vec{e}_z)$ is chosen so that $\vec{e}_x$, $\vec{e}_y$ and $\vec{e}_z$ are parallel to the crack propagation, tension loading at the crack tip and sample thickness directions respectively.

\section{Experimental results: Damage spreading mechanisms within the process zone}\label{sec_observation}

\subsection{Evidence of damage cavities ahead of the crack tip}

\begin{figure}[!htbp]
\centering
\includegraphics[width=0.9\textwidth]{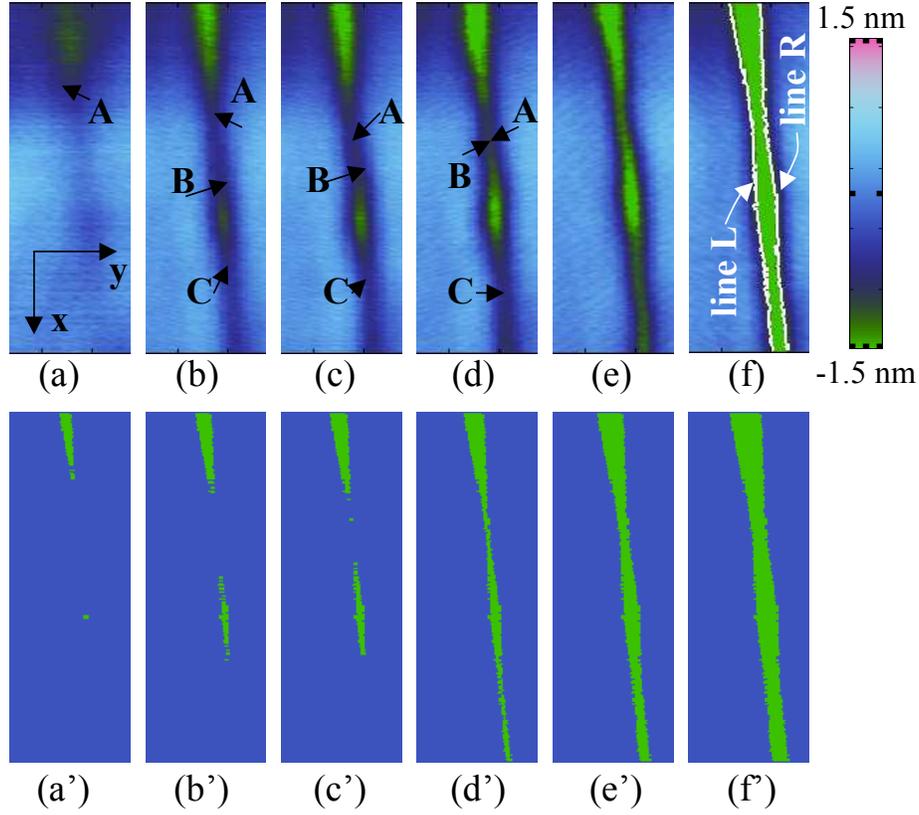}
\caption{(a)-(f) Sequence of topographic AFM frames in the vicinity of the crack tip
showing the propagation of the crack in stress corrosion regime ($v=4.10^{-11}\un{m.s}^{-1}$). The scan size is $470 \times 135\un{nm}^2$ and the height range over $3\un{nm}$. The 6 frames were taken during the five hours necessary for the main crack to cross the area of interest. (a,b) Appearance of a nanometric damage cavity ahead of the crack tip, (b-d) growth of the cavity prior to crack propagation and (f,g) coalescence of the cavity with the main crack. The points A, B and C locate the position of the main crack tip (CT), the backward front (BF) and the forward front (FF)
tips of the cavity respectively. (a')-(f') reconstructed frames using the FRASTA method}
\label{cavity}
\end{figure}

\begin{figure}[!htbp]
\centering
\includegraphics[width=0.9\textwidth]{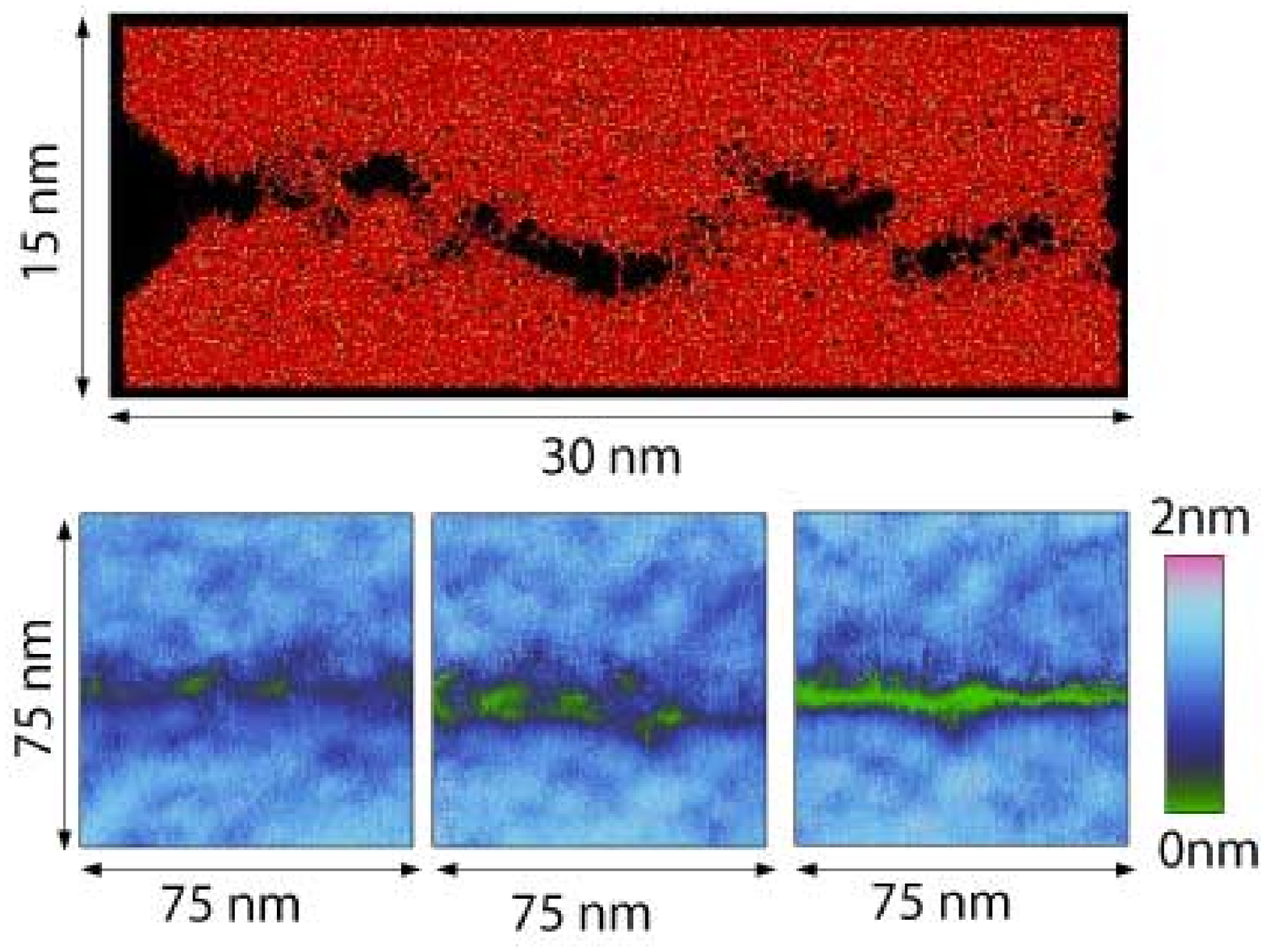}
\caption{Top: Snapshots of atoms as observed in a MD simulation of dynamic fracture in amorphous Silica (from Kalia {\em et al.}, 2003). Bottom: Sequences of three sucessive AFM snapshots showing the vicinity of the crack tip at the surface of an aluminosilicate glass specimen (from C{\'e}lari{\'e} {\em et al.}, 2003). In both cases, the crack progresses through the growth and coalescence of damage cavities}
\label{cavity_bib}
\end{figure}

Our experimental setup has allowed to observe the crack progression within the process zone in real time at the nanoscale. Figures \ref{cavity}(a-f) show six successive AFM topographic frames in the vicinity of the crack tip in the pure silica glass. For this specific sequence, the crack propagates at an average velocity $v \simeq 4.10^{-11}\un{m/s}$. One can clearly see a depression 
ahead of the crack tip. This cavity grows to a typical size of $100\un{nm}$ in length and $20\un{nm}$ in width, and then merges with the advancing main crack to make it cross the whole area of observation. In other words, the crack front does not propagate regularly as commonly stated (\opencite{Kelly67_pm}; \opencite{Lawn80_jms}; \opencite{Guin05_prl}), but progresses through the growth and coalescence of cavities. This scenario is fully consistent with what was observed both experimentally in aluminosilicate vitroceramic under stress corrosion by \citeauthor{Celarie03_prl} \shortcite{Celarie03_prl,Celarie03_ass} and \inlinecite{Marliere03_jpcm} (Figure \ref{cavity_bib}(bottom)) and numerically in dynamically breaking samples of amorphous Silica by \inlinecite{vanBrutzel02_mrssp}, \inlinecite{Rountree02_armr} and \inlinecite{Kalia03_ijf} (Figure \ref{cavity_bib}(top)). 

Fracture Surface Topography Analysis (\opencite{Kobayashi87_mt}; \opencite{Miyamoto90_ijf}) was then performed to ensure that the spots observed ahead the crack tip correspond actually to damage cavities. In a ductile scenario, the growth of damage cavities is expected to induce irreversible plastic deformations that will leave visible prints on the {\em post-mortem} fracture surfaces. We have thus determined the fracture lines after the crack has crossed the area of interest (white lines in Fig. \ref{cavity}f) and reconstituted virtually the intact material by placing the line on the right -~line R~- to the left of the line on the left -~line L \cite{Bonamy05_ijmpt}. By translating gradually the line R to the right, one reproduced qualitatively the chronology of the cavity growth (Fig. \ref{cavity}a'-f'). This provides a rather strong argument to relate these nano-scale spots to damage cavities.

It is worth to note that the reconstructed cavity is found to be thinner than the real one. This can be understood as follows: The shape of the real cavity is given not only by the irreversible -~permanent~- part of the opening displacements at the free surface, but also by the reversible -~elastic deformation~-. This latter vanishes once the crack has crossed the area of interest and the stresses have relaxed, - and therefore can not be taken into account in the reconstructed frames \cite{Bonamy05_ijmpt}. In others words the remnants of these nanospots growing ahead of the crack tip on the post-mortem fracture surfaces are expected to be much smaller - and thus much harder to detect~- than the spots themselves. This may explain why \inlinecite{Guin05_prl} did not succeed to see evidences of the pores' remnants from the analysis of the mismatch between the two opposite post-mortem fracture surfaces.

\subsection{Kinematics of damage cavities}\label{subsec_kinematic}

\begin{figure}[!htbp]
\centering
\includegraphics[height=0.4\textwidth]{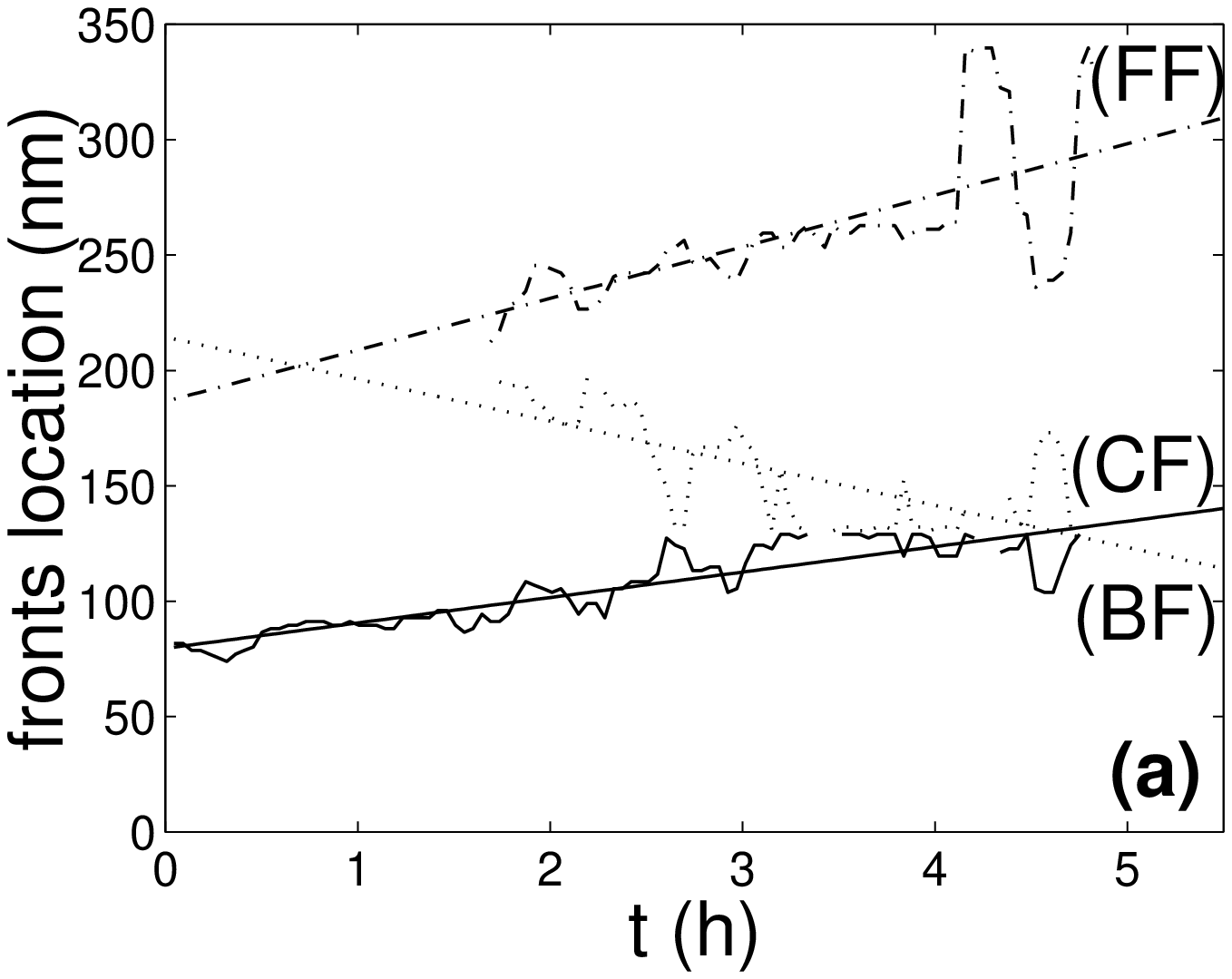}
\includegraphics[height=0.4\textwidth]{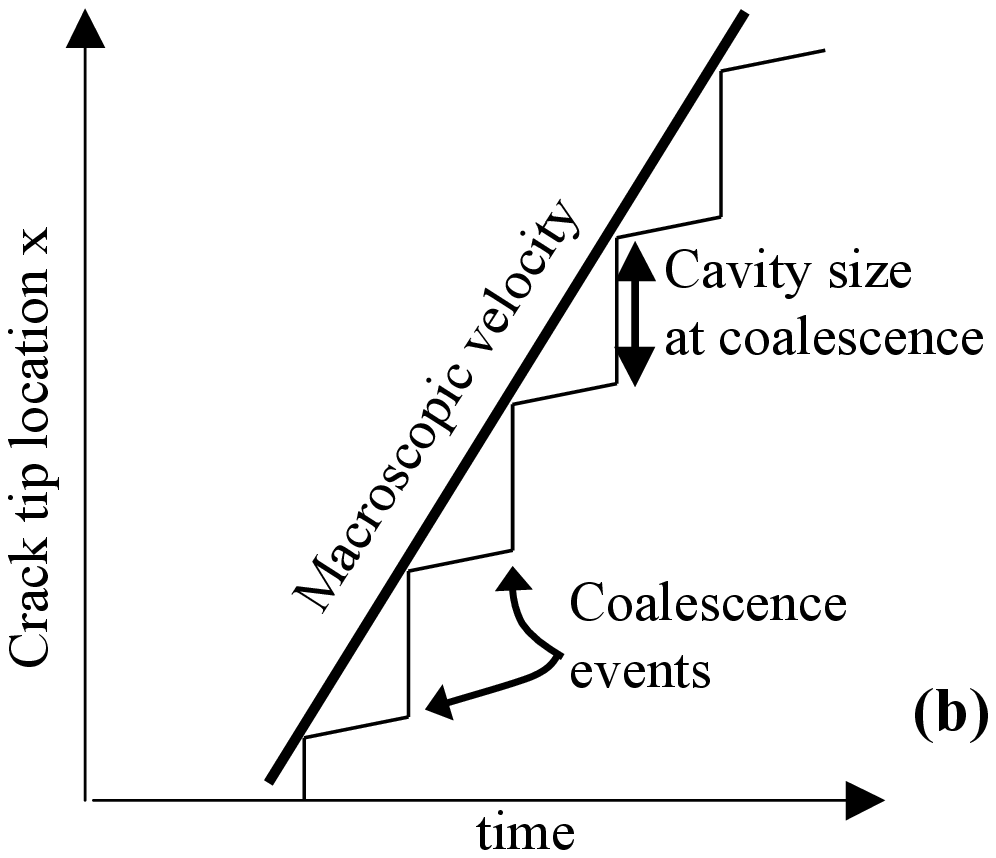}
\caption{(a) Temporal evolution of the main crack front (CF) and the backward front (BF) of the cavity and the forward front (FF) of the cavity
 represented by points $A$, $B$ and $C$ on the frames of Fig. \ref{cavity}(a)-(e) respectively. The velocities of these fronts are determined through linear fits: $v^{CF} \simeq 4.10^{-12}\un{m.s}^{-1}$, $v^{BF} \simeq 1.2.10^{-11}\un{m.s}^{-1}$ and $v^{FF} \simeq 1.1.10^{-11}\un{m.s}^{-1}$ . The thicker line shows the velocity at the continuum scale: $<v> \simeq 4.10^{-11}\un{m.s}^{-1}$. (b) Sketch of the crack tip propagation at the scale of the process zone.}
\label{kinematics}
\end{figure}

From the sequences represented in Figure \ref{setup}(a-f), the temporal evolutions of the main crack tip (CT), the forward front tip (FF) 
and the backward front tip (BF) of the cavity were then determined, and the corresponding velocities extracted (Figure \ref{kinematics}). 
These velocities were found to be $v^{CT}=4.10^{-12}\un{m/s}$, $v^{FF}=1.1.10^{-11}\un{m/s}$ and $v^{BF}=1.2.10^{-11}\un{m/s}$ for the main crack tip, the forward front tip and the backward front tip of the cavity, respectively, i.e. significantly smaller than the mean crack tip velocity $v=4.10^{-11}\un{m/s}$ as measured optically \cite{Prades05_ijss}. In other words, the crack growth velocity as observed at the continuum scale is dominated by the accelerating phases corresponding to cavity coalescence with the main crack front.

\subsection{Nanocavitation and process zone}\label{subsec_processzone}

\begin{figure}[!htbp]
\centering
\includegraphics[height=0.4\textwidth]{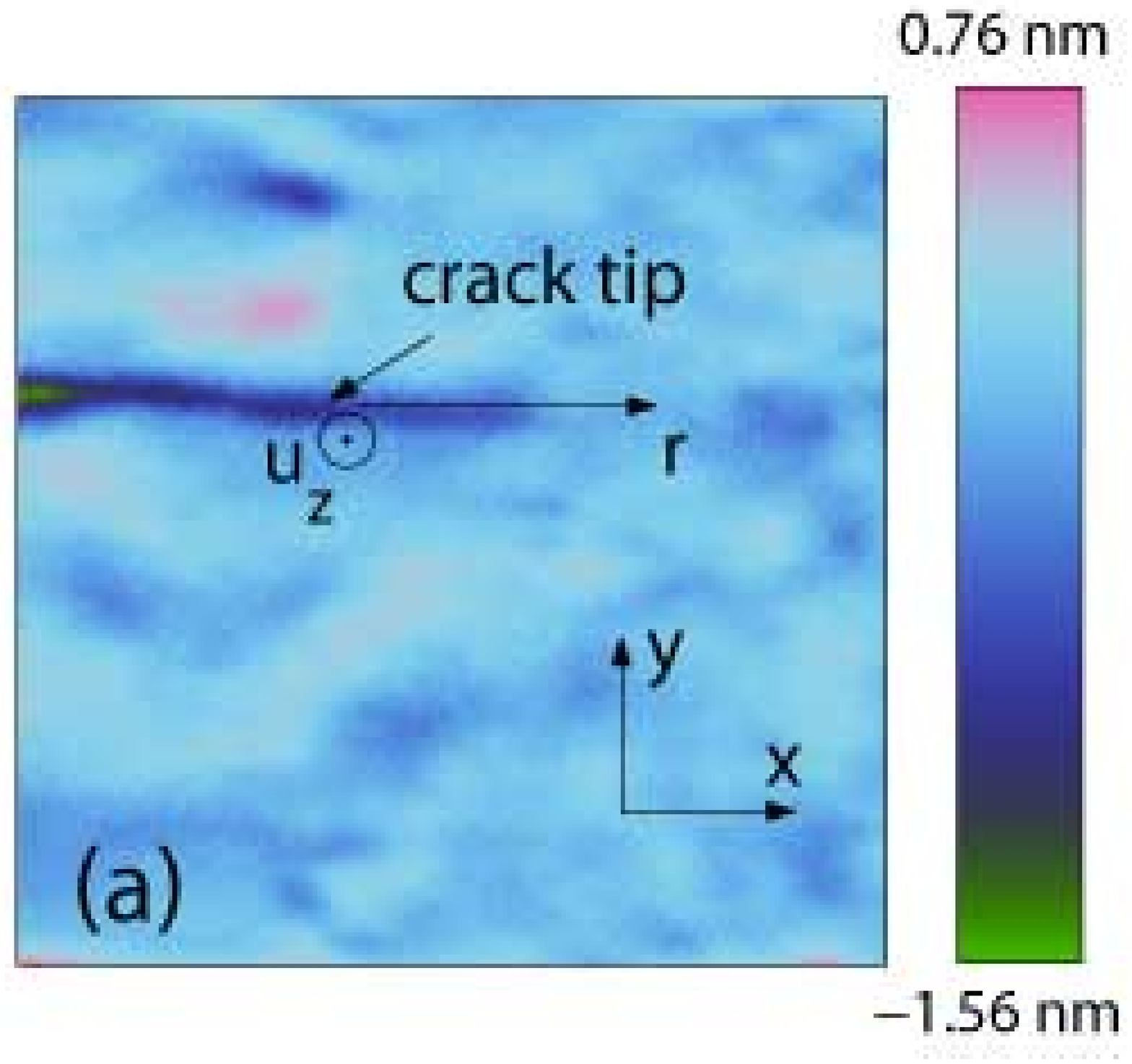}
\includegraphics[height=0.4\textwidth]{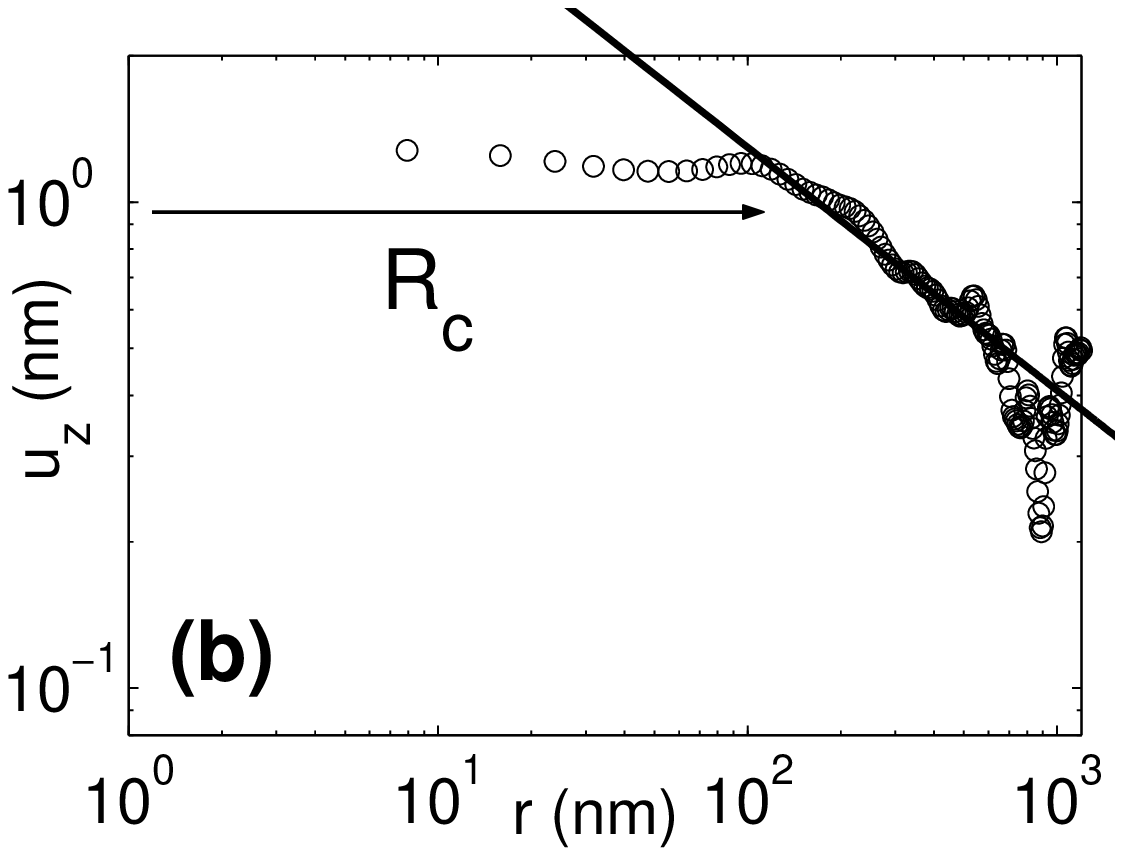}
\caption{Measurement of the extension of process zone. (a) $1\times1~\mu\mathrm{m}$ AFM topographical frame of the vicinity of the crack tip. The crack growth velocity was $v \simeq 3.10^{-11}\un{m/s}$ (b) Variation of the out-of-plane displacement $u_z$ as a function of the distance $r$ from the crack tip measured along the direction of crack propagation ($x$-axis). The axes are logarithmic. The straight red line corresponds to the LEFM predictions. For $r$ smaller than a given value $R_c \simeq 200\un{nm}$, the experimental curve departs from the LEFM predictions. This value sets the extent of the process zone along the $x$ direction for this specific experiment.}
\label{processzone}
\end{figure}

We then determined the extension of the process zone (\opencite{Bonamy05_ijmpt}; \opencite{Prades05_prl}). Linear Elastic Fracture Mechanics (LEFM) predicts that the stress field exhibits a universal square root singularity at the crack tip in an intact linear elastic medium. One thus expects a square root singularity in the components of the stress tensors, and in the out-of-plane displacement at the free surface of the specimen. The surface topography was thus imaged in the vicinity of the crack tip (Figure \ref{processzone}a), and compared to the LEFM predictions 
(Figure \ref{processzone}b). Far from the crack tip, one recovers the LEFM predictions. But below a given threshold, the experimental curves depart from the predictions. This departure point sets the extent of the process zone (\opencite{Guilloteau96_epl}; \opencite{Heneau00_jacs}; \opencite{ Celarie03_prl}). This extension was found to be around $100-300\un{nm}$ in length ($x$ direction) and around $20\un{nm}$ in width ($y$ direction) in the stress corrosion experiment ($v$ ranging from $8.10^{-12}$ to $10^{-9}\un{m/s}$) (\opencite{Bonamy05_ijmpt}; \opencite{Prades05_prl}). It was found to be significantly smaller, around $10-25\un{nm}$ in length, and $7-11\un{nm}$ in width for the MD simulation of dynamic crack propagation ($v$ ranging from $10$ to $400\un{m/s}$) \cite{Prades05_prl}.

We finally compared the extension of the process zone as defined above to the extent of the zone made "porous" in the vicinity of the crack due to the presence of cavities. Those were shown to coincide \cite{Prades05_prl}. This indicates that nucleation and growth of cavities provide the dominant mechanism responsible for damage spreading within the process zone. 

\section{Discussion: On the relevant lengthscales}\label{subsec_discussion}

Observations of damage cavities in amorphous silica under stress corrosion are consistent with what was observed experimentally in aluminosilicate vitroceramics under stress corrosion by \citeauthor{Celarie03_prl} \shortcite{Celarie03_prl,Celarie03_ass} and \inlinecite{Marliere03_jpcm},and numerically during dynamic failure of amorphous Silica by \inlinecite{vanBrutzel02_mrssp}, \inlinecite{Rountree02_armr}, and \inlinecite{Kalia03_ijf}. This indicates that the origin of such a nanoscale damage mode is inherent to the amorphous structure and does not depend on the precise glass composition. Origin of the damage cavities should be found in the amorphous structure, which contains {\em inherent density and/or stress fluctuations at the nanoscale}.

\subsection{Cavity size at coalescence}

In this scenario, the typical size of the damage cavities can be understood as follows. Intrinsic amorphous fluctuations induce low toughness regions with a density $\rho_0$ of weak bonds, \ie bonds that can be broken ahead of the crack tip. These "nano-cracks" behave as stress 
concentrators and can grow under the stress imposed by the presence of the main crack, to give birth to damage cavities. In dynamic fracture, the stress within the process zone is sufficient to activate any of these "low toughness sites". The typical distance between two nucleation sites - and consequently the typical size of cavities at coalescence - is given by the lengthscale $\delta_0$ of the fluctuations induced by the amorphous structure. Since $\delta_0$ is of the order of the nanometer, one expects a cavity size of the order of one nanometer which is compatible with the MD observations (\opencite{vanBrutzel02_mrssp}; \opencite{Rountree02_armr}; \opencite{Kalia03_ijf}). 

This scenario can be extended to the stress corrosion regime \cite{Prades05_prl}. In that case, the stress level in the process zone is too low to make low toughness sites give birth to damage cavities. However, the corrosive action of water activates some of them. The probability
 $\rho/\rho_0$ -~where $\rho$ refers to the density of activated sites~- is expected to scale as $\rho/\rho_0\propto \exp(\alpha G/kT)$ where $\alpha$ $G$, $k$ and $T$ refer to a typical activation area, the mechanical energy within the process zone, the Boltzmann constant, 
and the temperature respectively. Since the crack growth speed $v$ scales also as $v \propto \exp(\alpha G/kT)$ (\opencite{Wiederhorn67_jacs}; \opencite{Wiederhorn70_jacs}; \opencite{Prades05_ijss}), one gets $\rho/\rho_0 \propto v$. The typical distance $\delta$ between two nucleation points - and consequently the typical size of the cavity at coalescence - scales as $\delta\propto \rho^{-1/3}$. Hence $\delta/\delta_0 \propto v^{-1/3}$. In glasses under stress corrosion, crack growth stops being dominated by the rate of hydrolysis chemical reactions when $v$ is greater than a threshold $v^*$ classically described as the limit between stress corrosion
 regions I and II in the literature (\opencite{Wiederhorn67_jacs}; \opencite{Wiederhorn70_jacs}; \opencite{Lawn93_book}). The corresponding mechanical energy $G(v = v^*)$ is thus expected to be sufficient to activate all the low toughness sites: $\delta(v=v^*) \simeq \delta_0$. Finally, one gets $\delta/\delta_0 \simeq (v / v^*)^{-1/3}$ in the ultra-slow stress corrosion regimes.
 For our experimental conditions (relative humidity $45\%$, temperature $26^\circ \mathrm{C}$), $v^* \simeq 10^{-5}\un{m/s}$ (\opencite{Wiederhorn67_jacs}; \opencite{Wiederhorn70_jacs}; \opencite{Lawn93_book}). Since $\delta_0 \simeq 1 \un{nm}$, one gets $\delta \simeq 100\un{nm}$ 
for $v \simeq 4.10^{-11}\un{m/s}$, which is in good agreement with observations (Figure \ref{cavity}).

\subsection{Size of the process zone}\label{subsec_cavitysize}

Let us now discuss the typical size $R_c$ of the process zone. Analysis performed by \inlinecite{Swiler95_jncs} suggests the presence of voids as large as $0.45\un{nm}$ in unstressed silica glass. In first approximation, the process zone extent can be assimilated to the maximum distance from the crack tip where the stress imposed by the presence of the main crack is sufficient to make such a void grow. At this distance, the average stress level is expected to be $K_I/\sqrt{2\pi R_c}$ \cite{Inglis13_tina}. This stress is then concentrated at the notches of the voids and becomes $\sigma^{voids}\simeq (1+2r)K_I/\sqrt{2\pi R_c}$ where $r$ refers to the typical aspect ratio of the void. The void will grow when $\sigma^{voids}$ becomes larger than the intrinsic strength of the material $\sigma^*$. Hence, $R_c \simeq (1/2\pi)(1+2r)^2(K_I/\sigma^*)^2$.

The value of intrinsic strength of silica glass measured in vacuum at room temperature is estimated around $\sigma^* \simeq 10-12\un{GPa}$ \cite{Kurkjian03_jncs}. Voids in unstressed silica glass are expected to be around $0.2-0.45\un{nm}$ thick and $0.45\un{nm}$ large \cite{Swiler95_jncs}, which leads to an aspect ratio $r \simeq 1-2$. As dynamic fracture occurs for $K_I=K_{Ic}=1\un{MPa.m}^{1/2}$, one thus expects a process zone size $R_c\simeq 10 - 40\un{nm}$, which is in good agreement with the measurements performed on the MD simulations (section \ref{subsec_processzone}).

The intrinsic strength of Silica glass in humid air was found to be significantly smaller, around $\sigma^* \simeq 3-4\un{GPa}$ \cite{Proctor67_prsa}. Since the stress corrosion experiments were performed with a stress intensity factor $K_I \simeq 0.4-0.5$, one expects a process zone size $R_c \simeq 15-110\un{nm}$. This order of magnitude is compatible with what was measured from the 
AFM topography (Figure \ref{processzone}). One may however wonder how water molecules can travel from the crack tip along such a large distance within the glass. For water to travel over this distance in typically an hour, it requires a diffusion coefficient $D\simeq 10^{-14}\un{cm}^2\un{s}^{-1}$  \cite{Bonamy05_ijmpt} fives orders of magnitude larger than what is commonly observed in the absence of stress ($D\simeq 10^{-19}\un{cm}^2\un{s}^{-1}$). However, Tomozawa and Han \cite{Tomozawa91_jacs} have shown that under stress, diffusion is considerably accelerated, and hence it is perfectly plausible that water molecules react with Si-O bonds at a 
distance $\sim $100 nm from the crack tip within the bulk, under our experimental conditions.

\section{Conclusion}

Ultra-slow crack propagation was observed in real time at the nanoscale through AFM in amorphous Silica. The main results from our observations are:

\begin{itemize}
\item[(i)] Within the process zone, the crack tip does not propagate regularly, but through the growth and coalescence of damage cavities.

\item[(ii)] Velocity of the main crack tip and the cavity tips as measured at the damage zone scale are shown to be significantly smaller 
than the mean crack growth velocity as measured at the continuous scale. In other words, crack growth is intermittent and dominated by the accelerating phases corresponding to the cavity coalescence with the main crack front. 

\item[(iii)] The nanocavitation process provides the dominant mechanism responsible for damage spreading within the process zone. 

\end{itemize}

Observations of damage cavities in amorphous Silica under stress corrosion are consistent to what was observed experimentally in Aluminosilicate glasses under stress corrosion \citeauthor{Celarie03_prl} \shortcite{Celarie03_prl,Celarie03_ass} and \inlinecite{Marliere03_jpcm}, and numerically during dynamic failure of amorphous Silica (\opencite{vanBrutzel02_mrssp}; \opencite{Rountree02_armr}; \opencite{Kalia03_ijf}). This indicates that the existence of this nanoductile mode is inherent to the amorphous structure and does not depend on the precise glass composition.

The origin of cavitation should be found in the intrinsic toughness fluctuations induced by the amorphous structure of the material. This results in low toughness region sites and/or low density regions that behave as stress concentrators and give birth to cavities ahead of the main crack tip. Such a scenario was shown to capture quantitatively the cavity size at coalescence and the process zone extension in both the dynamic fracture simulations and the stress corrosion experiments.

This failure mechanism through growth and coalescence of damage cavities is very similar to what is observed classically - albeit at other length scales - in a wide range of materials, \eg  aluminosilicate vitroceramics (\citeauthor{Celarie03_prl}, \citeyear{Celarie03_prl,Celarie03_ass}; \opencite{Marliere03_jpcm}), nanophase materials (\opencite{Rountree02_armr}; \opencite{Kalia03_ijf}), PMMA \cite{Ravichandar97_jmps} and polymers \cite{Lapique02_japs}. We argue that such mechanism is generic to crack propagation - the main difference resides in the typical lengthscales over which cavities are observed. These lengthscales are controled by the typical size of the microstructure eventually modified through environmental assisted activated process like stress corrosion or fatigue. The aspect ratio of the cavities - much larger in glasses and quasi-brittle materials than in metallic alloys~- reflects the ability of the material to deform irreversibly. Such a scenario may explain the puzzling similarities observed in the scaling properties exhibited by fractures surfaces in a wide range of materials \cite{Bouchaud97_jpcm}.

Let us finally add that the present study sheds light on the role of the {\em spatial} fluctuations induced by the material microstructure, but passes over the {\em temporal} fluctuations in silence. Interaction of a growing crack with the material microstructure results in the release of acoustic waves (\opencite{Ravichandar84_ijf}; \opencite{Bonamy03_prl}) which play a significant role in the energy dissipation properties within the process zone. Their understanding represents interesting challenges for future investigations. 

\section*{Acknowledgements}

We acknowledge Thierry Bernard for his technical support. We are also indebted to Fabrice C{\'e}lari{\'e}, Stephane Chapuliot, Rajiv Kalia, Christian Marli{\`e}re, Laurent Van Brutzel, and Sheldon Wiederhorn for enlightning discussions, and to the MATCO programm for its constant support. CLR is currently supported by the National Science Foundation under Grant No. 0401467.

\end{article}


\begin{thebibliography}{}

\bibitem[\protect\citeauthoryear{Bonamy and Ravi-Chandar}{2003}]{Bonamy03_prl}
Bonamy D., and Ravi-Chandar K.: 2003, `Interaction of shear waves and propagating cracks'.
\newblock {\em Physical Review Letters} {\bf 91}, 235502-1--4.

\bibitem[\protect\citeauthoryear{Bonamy {\em et al.}}{2005}]{Bonamy05_ijmpt}
Bonamy D., Prades S., Ponson L., Dalmas D., Rountree C.~L., Bouchaud E. and Guillot C.: 2005, `Experimental Investigation of damage and Fracture in Glassy Materials at the nanometer scale'.
\newblock {\em submitted to International Journal of Materials and Product Technology}.

\bibitem[\protect\citeauthoryear{Bouchaud}{1997}]{Bouchaud97_jpcm}
Bouchaud E.: 1997, 'Scaling properties of cracks'.
\newblock {\em Journal of Physics (Condensed Matter)} {\bf 9}, 4319.

\bibitem[\protect\citeauthoryear{C{\'e}lari{\'e} {\em et al.}}{2003a}]{Celarie03_prl}
C{\'e}lari{\'e} F., Prades S., Bonamy D., Ferrero L., Bouchaud E., Guillot C., Marli{\`e}re C.: 2003, Glass breaks like metal, but at the nanometer scale, Glass breaks like metal, but at the nanometer scale'.
\newblock {\em Physical Review Letters} {\bf 90}, 075504/1-4.

\bibitem[\protect\citeauthoryear{C{\'e}lari{\'e} {\em et al.}}{2003b}]{Celarie03_ass}
C{\'e}lari{\'e} F., Prades S., Bonamy D., Dickel{\'e} A., Bouchaud E., Guillot C., Marli{\`e}re C.: 2003, 'Surface fracture of glassy materials as detected by real-time atomic force microscopy (AFM) experiments'.
\newblock {\em Applied Surface Science} {\bf 212}, 92-96.

\bibitem[\protect\citeauthoryear{Guilloteau {\em et al.}}{1996}]{Guilloteau96_epl}
Guilloteau E., Charrue H., Creuzet F.: 1996. 'The direct observation of the core region of a propagating fracture crack in glass.'
\newblock {\em Europhysics Letters} {\bf 34}, 549-553.

\bibitem[\protect\citeauthoryear{Guin and Wiederhorn}{2005}]{Guin05_prl}
Guin J.~P. and Wiederhorn S.~M.: 2005. 'Fracture of Silicate Glasses: Ductile or Brittle?'
\newblock {\em Physical Review Letters} {\bf 92}, 215502/1-4.

\bibitem[\protect\citeauthoryear{He {\em et al.}}{1995}]{He95_amm}
He M.Y., Turner M.~R. and Evans A.~G.: 1995. 'Analysis of the double cleavage drilled compression specimen for interface fracture energy measurements over range of mode mixities'.
\newblock {\em Acta Metallurgica Materialia} {\bf 43} 3453-3458.

\bibitem[\protect\citeauthoryear{H{\'e}naux and Creuzet}{2000}]{Heneau00_jacs}
Henaux S., Creuzet F.: 2000. 'Crack tip morphology of slowly growing cracks in glass.'
\newblock {\em Journal of the American Ceramic Society} {\bf 83} 415-417.

\bibitem[\protect\citeauthoryear{Kelly {\em et al.}}{1967}]{Kelly67_pm}
Kelly A., Tyson W.~R., Cottrell A.~H.: 1967. 'Ductile and brittle crystals.'
\newblock {\em The Philosophical Magazine} {\bf 15} 567--586.

\bibitem[\protect\citeauthoryear{Kobayashi and Shockey}{1987}]{Kobayashi87_mt}
Kobayashi T. and Shockey D.A.: 1987. 'A Fractographic Investigation of Thermal Embrittlement in Cast Duplex Stainless Steel.'
\newblock {\em Metallalurgical Transactions A} {\bf 18A} 1941-1949.

\bibitem[\protect\citeauthoryear{Kalia {\em et al.}}{2003}]{Kalia03_ijf}
Kalia R.~., Nakano A., Vashishta P., Rountree C.~L., van Brutzel L., Ogata, S.: 2003. 'Multiresolution atomistic simulations of dynamic fracture in nanostructured ceramics and glasses.'
\newblock {\em International Journal of Fracture} {\bf 18A} 1941-1949.

\bibitem[\protect\citeauthoryear{Kurkjian {\em et al.}}{2003}]{Kurkjian03_jncs}
Kurkjian C.~R., Gupta P.~K., Brow R.~K., Lower N.: 2003. 'The intrinsic strength and fatigue of oxide glasses.'
\newblock {\em Journal of Non-Crystalline Solids} {\bf 316} 114-124.

\bibitem[\protect\citeauthoryear{Inglis}{1913}]{Inglis13_tina}
Inglis C.~E.: 1913. 'Stresses in a plate due to the presence of cracks and sharp corners.'
\newblock {\em Transactions of the Institution of Naval Architects} {\bf 55} 219--230.

\bibitem[\protect\citeauthoryear{Lapique {\em et al.}}{2002}]{Lapique02_japs}
Lapique F., Meakin P., Feder J., Jossang T.: 2002. '.'
\newblock {\em Journal of Applied Polymer Sciences} {\bf 86} 973--983.

\bibitem[\protect\citeauthoryear{Lawn}{1993}]{Lawn93_book}
Lawn B.~R.: 1993.
\newblock {\em Fracture of Brittle Solids}[2nd edition] Cambridge: Cambridge University Press, 

\bibitem[\protect\citeauthoryear{Lawn {\em et al.}}{1980}]{Lawn80_jms}
Lawn B.~R., Hockey B.~J. and Wiederhorn S.~M.: 1980. 'Atomically sharp cracks in brittle solids: an electron microscopy study.'
\newblock {\em Journal of Materials Science.} {\bf 15} 12.

\bibitem[\protect\citeauthoryear{Marli{\`e}re {\em et al.}}{2003}]{Marliere03_jpcm}
Marli{\`e}re C., Prades S., C{\'e}lari{\'e} F., Dalmas D., Bonamy D., Guillot C., Bouchaud E.: 2003. 'Crack fronts and damage in glass at the nanometre scale.'
\newblock {\em Journal of Physics-Condensed Matter} {\bf 15} 2377-2386. 

\bibitem[\protect\citeauthoryear{Miyamoto {\em et al.}}{1990}]{Miyamoto90_ijf}
Miyamoto H., Kikuchi M., Kawazoe T.: 1990. 'A study on the ductile fracture of Al-alloys 7075 and 2017.'
\newblock {\em International Journal of Fracture} {\bf 42} 389-404. 

\bibitem[\protect\citeauthoryear{Paun and Bouchaud}{2003}]{Paun03_ijf}
Paun F., Bouchaud E.: 2003, 'Morphology of damage cavities in aluminium alloys'.
\newblock {\em International Journal of Fracture} {\bf 121}, 43--54.

\bibitem[\protect\citeauthoryear{Prades {\em et al.}}{2005}]{Prades05_ijss}
Prades S., Bonamy D., Dalmas D., Bouchaud E. and Guillot C.: 2005, `Nano-ductile crack propagation in glasses under stress corrosion: spatiotemporal evolution of damage in the vicinity of the crack tip'.
\newblock {\em International Journal of Solids and Structures } {\bf 42}, 637-–645.

\bibitem[\protect\citeauthoryear{Prades {\em et al.}}{2005}]{Prades05_prl}
Prades S., Rountree, C.~L., Bonamy D., Dalmas D., Bouchaud E., Kalia R.~K., Guillot C.: 2005, 'A unified damage scenario in glasses: from ultraslow to ultrafast fracture'.
\newblock {\em submitted to Physical Review Letter}.

\bibitem[\protect\citeauthoryear{Proctor {\em et al.}}{1967}]{Proctor67_prsa}
Proctor B.~A., Whitney I., johnson J.~W.: 1967, 'The strength of fused
silica,'.
\newblock {\em Proceedings of the Royal Society} {\bf A297}, 534--557.

\bibitem[\protect\citeauthoryear{Ravi-Chandar and Knauss}{1984}]{Ravichandar84_ijf}
Ravi-Chandar K. and Knauss W.~G.: 1984, 'An experimental investigation into dynamic fracture-IV On the interaction of stress waves with propagating cracks'.
\newblock {\em International Journal of Fracture} {\bf 26}, 189--200.

\bibitem[\protect\citeauthoryear{Ravi-Chandar and Yang}{1997}]{Ravichandar97_jmps}
Ravi-Chandar K. and Yang B.: 1997, 'On the role of microcracks in the dynamic fracture of brittle materials'.
\newblock {\em Journal of the Mechanics and Physics of Solids} {\bf 45}, 535--563.

\bibitem[\protect\citeauthoryear{Rountree {\em et al.}}{2002}]{Rountree02_armr}
Rountree C.~L., Kalia R.~K., Lidorikis E., Nakano A., Van Brutzel L., Vashishta P.: (2002). 'Atomistic aspects of crack propagation in brittle materials: Multimillion atom molecular dynamics simulations.'
\newblock {\em Annual Review of Material Research} {\bf 32}, 377--400.

\bibitem[\protect\citeauthoryear{Swiler {\em et al.}}{1995}]{Swiler95_jncs}
Swiler T.~P., Simmons J.~H., Wright A.~C.: 1995, 'Molecular dynamics study of brittle fracture in silica glass and cristobalite.'
\newblock {\em Journal of Non-Crystalline Solids} {\bf 182}, 68--77.

\bibitem[\protect\citeauthoryear{Tomozawa {\em et al.}}{1991}]{Tomozawa91_jacs}
Tomozawa M., Han W.~H., Lanford W.~A.: 1991, 'Water Entry into Silica Glass During Slow Crack Growth.'
\newblock {\em Journal of the American Ceramic Society} {\bf 74}, 2573-2576.

\bibitem[\protect\citeauthoryear{Van Brutzel {\em et al.}}{2002}]{vanBrutzel02_mrssp}
Van Brutzel L., Rountree C.~L., Kalia R.~K., Nakano A., Vashishta P.: (2002). 'Dynamic fracture mechanisms in nanostructured and amorphous silica glasses: million-atom molecular dynamics simulations.'
\newblock {\em Materials Research Society Symposium Proceedings} {\bf 703}, V.3.9.1--V.3.9.6.

\bibitem[\protect\citeauthoryear{Wiederhorn}{1967}]{Wiederhorn67_jacs}
Wiederhorn S.~M.: 1967, 'Influence of water vapor on crack propagation in Soda-lime glass.'
\newblock {\em Journal of the American Ceramic Society} {\bf 50}, 407-414.

\bibitem[\protect\citeauthoryear{Wiederhorn and Boltz}{1970}]{Wiederhorn70_jacs}
Wiederhorn S.~M. and Boltz L.~H.: 1970, 'Stress corrosion and static fatigue of glass.'
\newblock {\em Journal of the American Ceramic Society} {\bf 53}, 543-548.


\end{thebibliography}
\end{document}